\documentclass[aip,jmp,notitlepage,floatfix,amsmath,amssymb,
 reprint]{revtex4-2}

\usepackage{graphicx,color}
\usepackage{marvosym}
\usepackage{amsmath,amssymb}
\usepackage{vmargin}
\usepackage{graphicx,amsmath,graphics}
\usepackage{natbib} 
\usepackage{bm}
\usepackage{bbm}
\usepackage{dcolumn}   % needed for some tables
\usepackage[normalem]{ulem} % to use cross out text

\setmarginsrb{2cm}{2.cm}{2.cm}{1.5cm}%
             {11pt}{18pt}{11pt}{18pt}
\newcommand{\atomica}{\affiliation{Instituto Carlos I de F\'{\i}sica Te\'orica y Computacional and
      Departamento de F\'{\i}sica At\'omica, Molecular y Nuclear, Universidad de Granada, 18071
      Granada, Spain}}%
\newcommand{\mataplicada}{\affiliation{Departamento de Matem\'atica Aplicada, Universidad de Granada, 18071 Granada, Spain}}%
\newcommand{\uam}{\affiliation{Departamento de Qu\'imica, Universidad Aut\'onoma de Madrid, M\'odulo 13, Madrid, 28049, Spain}}
\newcommand{\ucm}{\affiliation{Departamento de Qu\'imica F\'isica, Universidad Complutense de Madrid, 28040 Madrid, Spain}}
\newcommand{\itamp}{\affiliation{ITAMP, Center for Astrophysics $\mid$ Harvard \& Smithsonian, Cambridge, Massachusetts 02138, USA}}

\newcommand{\ie}{i.\,e.}%
\newcommand{\gt}{>}
\newcommand{\lt}{<}

\usepackage[colorinlistoftodos]{todonotes}
\usepackage[colorlinks=true, allcolors=blue]{hyperref}

\renewcommand{\equationautorefname}{Eq.}
\def\equationautorefname~#1\null{Eq. (#1)\null}

\begin{document}

\title{An introduction to classical monodromy: applications to molecules in external fields  
%A semi-theoretical method to detect classical monodromy
}
\author{Juan J.\ Omiste}\ucm\uam
\author{Rosario Gonz\'alez-F\'erez}\atomica\itamp
\author{Rafael Ortega}\mataplicada

\date{\today}
\begin{abstract} 
An integrable Hamiltonian system presents monodromy if  the action-angle variables cannot  be defined globally. As a prototype of classical monodromy with azimuthal symmetry, we consider a linear molecule interacting with external fields and explore the topology structure of its phase space. 
Based on the behavior of closed orbits around
singular points or regions of the energy-momentum plane, 
a semi-theoretical method is derived to detect classical monodromy. The
validity of the monodromy test is numerically illustrated for several
systems with azimuthal symmetry. 
\end{abstract}
\pacs{}
\maketitle

\section{Introduction}
\label{sec:introduction}

Classical monodromy is the topological obstruction to define a global set of action-angle 
variables in certain integrable classical systems~\cite{Duistermaat1980}. 
This phenomenon has crossed the frontiers of mathematics, where it was  introduced by 
Duistermaat~\cite{Duistermaat1980} and initially considered as a mathematical curiosity with no 
significant physical applications, to acquire an interdisciplinary interest and popularity in 
classical and quantum physics. 
The quantum analog, \ie, quantum monodromy, is the impossibility of assigning a unique set 
of quantum numbers to characterize all states of a quantum 
system~\cite{Duistermaat1988,Zhilinskii2010}. 
The existence of monodromy has been proved, both theoretically and experimentally, 
for a wide variety of classical and quantum systems.

In classical mechanics, the action-angle variables characterize the dynamics of an
integrable system and determine the trajectories. Let us highlight that these quantities depend on the topological structure and properties of the phase space. 
The static  manifestation of monodromy, \ie,  the absence of global action-angle variables, 
is a singular fiber in the image of the energy-momentum
map as an isolated singular value~\cite{Duistermaat1980}.
The implications of monodromy have been extensively investigated for many classical systems, 
 such as the classical spherical pendulum~\cite{Duistermaat1988,Zhilinskii2010} or the
champagne bottle potential~\cite{Bates1991}, among many others. 
The classical monodromy has been experimentally investigated for the 1:1:2 resonant elastic 
pendulum showing  that  its  precession is a multivalued function of the constants of 
motion~\cite{fitch:phys_rev_lett_103}. 
In addition to these static manifestations, the monodromy presents dynamical consequences 
when a system is continuously  driven around a monodromy circuit~\cite{Delos2008,Chen2014}. 
This dynamical monodromy has been experimentally observed  as topological changes on the 
time-dependent evolution of a spherical pendulum driven by magnetic potentials~\cite{Nerem2018}. 

Quantum monodromy appears as a defect in the 
quantum lattice formed in the energy-momentum by the quantum eigenenergies~\cite{Duistermaat1988,Sadovskii2006,Zhilinskii2010}. 
As a consequence, the quantum monodromy significantly influences the spectra of atoms and 
molecules in external fields~\cite{schleif:pra2007,Kozin2003,Arango2004}, and even 
the stability of condensed bosons in optical lattices~\cite{Arwas2019}.
Quantum monodromy has been also encountered in the bending spectra of 
molecules~\cite{Child1999,Joyeux2003,Cushman2004}. For instance, the bending and 
symmetric stretching vibrations of the CO$_2$ molecule~\cite{Cushman2004}
provide a molecular realization of the quantum 1:1:2 resonant swing spring~\cite{Giacobbe2004}. 
Furthermore, quantum monodromy has been experimentally confirmed  in the energy-momentum 
maps of the  end-over-end rotational energy and the two-dimensional bending vibrational energy of 
cyanogen isothiocyanate molecules~\cite{Winnewisser2005} and in the bending levels of water 
molecules~\cite{Zobov2005}.
A dynamical manifestation of quantum monodromy has been theoretically explored in terms of
topological changes produced in the quantum wave function of the Mexican hat 
system~\cite{Chen2018}.

Classical monodromy is nowadays a well developed mathematical theory. For two-degree of freedom integrable systems, there are  several ways to detect monodromy, for instance, the most recent criteria are linked to the notion of focus-focus singularities~\cite{Martynchuk2021}.
%\old{and we refer to the recent survey~\cite{Martynchuk2021} for
%information on the most recent criteria linked to the notion of focus-focus singularity.}
However, the general reader can find some difficulties to access this theory since it requires rather sophisticated mathematical tools. The purpose of this paper is to review the origins of the theory and to make it accessible for the non-specialist, aiming at an audience of physicists, theoretical chemists or general dynamicists.

In this work, we present a semi-theoretical method combining an intuitive description of the family of invariant tori with the numerical computation of an integral. This integral, associated to a winding number, can only take discrete values, therefore, an approximate numerical computation allows to obtain the exact value. With these elementary tools, we design a method that could be considered insufficient from a strict mathematical point of view,  but certainly is convincing as a monodromy test. Specifically, we explore the topological features of classical monodromy and apply this monodromy test to a particle whose motion is constrained to the unit sphere
and governed by  a  potential with azimuthal symmetry. 
Following pioneering works on classical monodromy~\cite{Duistermaat1980,Duistermaat1988,Bates1991},
we study the behavior of closed orbits around 
singular points or regions of the corresponding energy-momentum plane, and present 
a monodromy test to identify and  characterize numerically the 
monodromy of this system.
The validity of this test, which is based on the definition of monodromy~\cite{Duistermaat1980,Duistermaat1988},
is illustrated for the unperturbed and  
perturbed 
classical spherical pendula, whose energy-momentum maps are also described.

The analyzed potentials also characterize the rotational dynamics of  a
linear molecule either in a static electric field~\cite{meyenn1970}, or in combined 
 static electric and non-resonant laser fields~\cite{friedrich:jcp111,nielsen:prl2012}, 
where the phenomenon of quantum monodromy has been previously analyzed~\cite{Arango2004}. 
This numerical study shows that the monodromy test  is easily implemented for different potentials 
and efficiently characterizes the monodromy of a wide variety of systems.

This work is organized as follows: in Sec.~\ref{sec:system_and_topology} we describe the dynamical systems under
study as well as the topology of the associated space. The phenomenon of monodromy is mathematically analyzed
in Sec.~\ref{sec:monodromy}, and an analytical test to determine if a dynamical system presents monodromy is derived. 
The validity of this monodromy test is illustrated numerically for several systems  with azymuthal symmetry  in Sec.~\ref{sec:numerical_tests}.
In the appendices we explain some details of the methods and theory discussed in the main text.
The main conclusions and perspectives of this work are presented in Sec.~\ref{sec:conclusions}.

\section{The system and its topology}
%\subsection{A generalized spherical pendulum}
\label{sec:system_and_topology}
The system is formed by a particle of mass $m=1$, whose motion  is constrained to the unit sphere 
\begin{equation*}
  M=T\mathbb{S}^2=\{(x,p)\in\mathbb{R}^3\times\mathbb{R}^3:\parallel x\parallel=1,\langle x,p\rangle =0\}
\end{equation*}
where $x$ and $p$ stand for the spatial coordinates and the momentum, respectively. 
The motion of the particle can be described using  Lagrangian coordinates  $\theta\in ]0,\pi[$ 
and  $\varphi\equiv \varphi+2\pi$, see Appendix~\ref{sec:symplectic} for more
details. 
As indicated  in this appendix, this chart does not cover the tangent planes 
at the north  and south poles at $\theta=0$ and $\theta=\pi$, respectively.
The particle interacts with a conservative force whose potential is invariant under rotations around the 
vertical  axis, \ie,  $V(\theta,\varphi)=V(\theta)$. 
This potential  is an analytic, non-constant, even, and $2\pi$-periodic function, \ie,
%\begin{equation*}
  %V\ne \text{constant},\quad 
 $ V(-\theta)=V(\theta)$ and $V(\theta+2\pi)=V(\theta)$.
%\end{equation*}
Note that throughout this work all functions (in one or several variables) are real analytic.
This type of potentials describe the interaction of a polar linear molecule 
 with external electric fields parallel to $Z$-axis on the laboratory fixed frame.
  For instance, the potential describing the interaction of a polar molecule with a static electric field~\cite{meyenn1970} is
 $V(\theta)=\epsilon_1\cos\theta$, which is the same potential as the classical spherical 
 pendulum. After adding a non-resonant laser to the static electric field the new potential is~\cite{friedrich:jcp111,nielsen:prl2012} 
$V(\theta)=\epsilon_1\cos\theta+\epsilon_2\cos^2\theta$. Finally, with a two-color non-resonant
laser field the potential becomes~\cite{oda2010} $V(\theta)=\epsilon_1\cos\theta+\epsilon_2\cos^2\theta+\epsilon_3\cos^3\theta$,
with $\epsilon_i$, $i=1,2,3$, being determined by the applied field strengths and the molecular features.

The Hamiltonian of this system reads
\begin{equation*}
  H=\frac{1}{2}\left(p_\theta^2+\cfrac{p_\varphi^2}{\sin^2\theta}\right)+V(\theta),
\end{equation*}
and the equations of motion are 
\begin{equation}
  \label{eq:em}
  \left\{
  \begin{array}{l}
\dot\theta=p_\theta,\quad
\dot p_\theta=\cfrac{p_\varphi^2\cos\theta}{\sin^3\theta}-V^\prime(\theta),\\
\dot\varphi=\cfrac{p_\varphi}{\sin^2\theta},\quad \dot p_\varphi=0.
\end{array}
\right.
\end{equation}
Note that due to the azimuthal symmetry,  $p_\varphi$ is an integral of motion. 
By fixing the energy $H=h$, and
the momentum $p_\varphi=j$, the variables $\theta$ and $\varphi$ satisfy
\begin{eqnarray}
  \label{eq:re}
  &&\frac{1}{2}\dot\theta^2+V_j(\theta)=h,\\
       \label{eq:re1}
   &&\dot\varphi=\cfrac{j}{\sin^2\theta}
\end{eqnarray}
where $V_j(\theta)$ is the modified potential
\begin{equation*}
  V_j(\theta)=\cfrac{j^2}{2\sin^2\theta}+V(\theta), 
\end{equation*}
with $V_0(\theta)=V(\theta)$.
The energy and momentum are first integrals in 
involution (in two degrees of freedom this  condition is always satisfied for first integrals), \ie, $  \{H,p_\varphi\}_M=0$, 
and  the Hamiltonian system is Liouville integrable on any region $\Omega\subset M$,
satisfying:
i) $\Omega$ is invariant under the flow~\eqref{eq:em}; and 
ii) the differentials $\mathrm{d}H$ and $\mathrm{d}p_\varphi$ are linearly independent on each
  point of $\Omega$~\cite{Arnold1989,Moser2006}.
These differentials, which are linear forms on the tangent space at each point of $M$, 
can be expressed
in terms of the basis $\{\mathrm{d}\theta, \mathrm{d}p_\theta, \mathrm{d}\varphi,\mathrm{d}p_\varphi\}$. The linear independence of $\mathrm{d}H$ and $\mathrm{d}p_\varphi$ is equivalent
to saying that the matrix
\begin{equation}
  \label{eq:sim}
  \cfrac{\partial(H,p_\varphi)}{\partial(\theta,p_\theta,\varphi,p_\varphi)}=
  \begin{pmatrix}
       V^\prime(\theta) -\frac{p_\varphi^2\cos\theta}{\sin^3\theta}& p_\theta & 0 & 
       \frac{p_\varphi}{\sin^2\theta}\\
    0 & 0& 0& 1
  \end{pmatrix}
\end{equation}
has rank two.
This is satisfied everywhere, except at the points fulfilling 
\begin{equation}
\label{eq:singularidad}
  p_\theta=0, \quad V^\prime(\theta)-\frac{p_\varphi^2\cos\theta}{\sin^3\theta}=0,
%  p_\theta=0\quad\text{and}\quad V_{p_\varphi}^\prime(\theta)=0.
%%% RGF  V_{p_\varphi}  no está definidida debería de ser  V_{j}
%%% JJO Cierto, deberíamos restringir todo a V_{j}
\end{equation}
where the rank is 1.

\subsection{Invariant tori}
\label{sec:inv}
\begin{figure*}[t]
  \centering
  \includegraphics[scale=1.2]{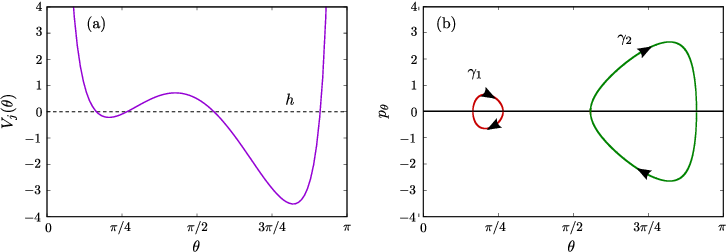}
\caption{(a) Modified potential $V_j(\theta)= 2\cos \theta-5\cos^2 \theta+j^2/(2\sin^2\theta)$ with $j=1$ as a function of $\theta$, the 
horizontal line represents $h=0$, 
(b) trajectories  $\gamma_1$ and $\gamma_2$ for $h=0$ in the phase space $(\theta,\,p_\theta)$.}
\label{fig:fig1}
\end{figure*}

\begin{figure*}[b]
  \centering
  \includegraphics[scale=0.8]{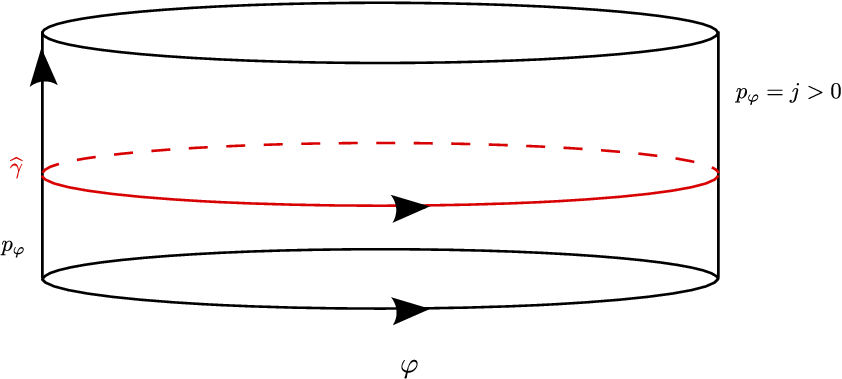}
  \caption{Phase space $(\varphi,j)$ for $j>0$ corresponding to different trajectories $\widehat{\gamma}$.}
  \label{fig:fig2}
\end{figure*}

The geometrical and dynamical structure of the sets 
${\cal S}_{h,j}=\{ (\theta,p_\theta,\varphi,p_\varphi):\, H=h,\, p_\varphi=j\}$ depends on the values of $j$ and $h$. 
For $j\ne0$,  the modified potential $V_j(\theta)$ 
diverges as $\theta\rightarrow 0^+$ or $\theta\rightarrow\pi^-$, \ie,
$ V_j(\theta)\rightarrow\infty$ as $\theta\rightarrow 0^+$ or $\theta\rightarrow\pi^-.$
For  a regular value $h>\min_{]0,\pi[}V_j(\theta)$,
\ie, $V_j^{\prime}(\theta)\ne 0$ for each $\theta$ such that $V_j(\theta)=h$, 
Eqs.~\eqref{eq:re} and~\eqref{eq:re1} 
describe a finite set of closed trajectories $\gamma_k$, with $k=1,\dots,r$, in the plane
$(\theta,p_\theta)$. 
An example is presented in Fig.~\ref{fig:fig1}~(a) with a prototype potential $V_j(\theta)$, 
with two closed trajectories, $r=2$, at $h=0$.
 For this potential and $h>\min_{]0,\pi[}V_j(\theta)$,
there are four turning points, as illustrated in Fig.~\ref{fig:fig1}.
An example of the trajectories in the phase space  are plotted in Fig.~\ref{fig:fig1}~(b). 
%% Primera figura de pagina 5.
The orbits $\gamma_k$ are periodic with   period 
\begin{equation}
  \label{eq:f1}
    T_k(h,j)
  =\sqrt{2}\int_{\alpha_-}^{\alpha_+}\cfrac{\mathrm{d}\theta}{\sqrt{h-V_j(\theta)}},
\end{equation}
where $\alpha_\pm=\alpha_\pm(h,j)$ are the intersection points of the orbit
$\gamma_k$ with the horizontal line $p_\theta=0$, see Fig.~\ref{fig:fig1}~(b).
In the cylinder $(\varphi,p_\varphi)$, the conservation of 
the angular momentum $p_\varphi$ 
and Eq.~\eqref{eq:re1} lead to a closed orbit $\widehat{\gamma}=\{(\varphi,j):\varphi\in[0,2\pi]\}$, 
an example is given in Fig.~\ref{fig:fig2}. 
For this orbit, the period 
$\widehat{T}_k=\widehat{T}_k(h,j)$ is implicitly defined by
\begin{equation}
  \label{eq:f2}
 j\int_0^{\widehat{T}_k}\cfrac{\mathrm{d}t}{\sin^2\theta(t)}=2\pi,
\end{equation}
and its orientation is determined by the sign of $j$.
%% Primera figura de pagina 6
Thus, %in the phase space ($\theta,p_\theta,\varphi,p_\varphi$), 
the set 
%$H=h$,$p_\varphi=j$ 
${\cal S}_{h,j}$ is composed by a
finite family of invariant tori $\gamma_k\times\widehat{\gamma},\,k=1,\ldots r$. When the periods 
$T_k$ and $\widehat{T}_k$ are commensurable,  
for instance, $ T_k=2\sqrt{2}$ and $\widehat{T}_k=3\sqrt{2}$,  then $\theta(t)$ and
$\varphi(t)$ have a common period and the invariant torus is foliated with closed orbits. 
If $T_k$ and $\widehat{T}_k$ are not commensurable,
 e.g., $ T_k=2\sqrt{2}$ and $\widehat{T}_k=3$,  the
orbits lying on the torus are quasiperiodic with frequencies
${2\pi}/{T_k},$ and ${2\pi}/{\widehat{T}_k}$.
%$\omega_1=\cfrac{2\pi}{T_k},$ and $\omega_2=\cfrac{2\pi}{\widehat{T}_k}$.

%It  has been assumed that $h$ is a regular value of $V_j(\theta)$ and this is essential in the previous discussion.

For $j\ne0$ and $h$ being  a critical value of $V_j$,  the set ${\cal S}_{h,j}$ 
%$\{H=h,p_\varphi=j\}$ 
can present
different topological configurations. 
For instance, the potential $V_j$  in Fig.~\ref{fig:fig1} has three critical values $h_1<h_2<h_3$,
corresponding to three equilibria $e_1$, $e_2$ and $e_3$, respectively. 
At the minimum $h=h_1$, ${\cal S}_{h,j}=\{e_1\}\times \hat{\gamma}$
%$\{H=h_1,p_\varphi=j\}$ 
is homeomorphic to $\mathbb{S}^1$, and at the local minimum $h=h_2$,
${\cal S}_{h,j}=(\gamma\times \{e_2\})\times \hat{\gamma}$, with $\gamma$ a closed orbit, 
  it is
homeomorphic to the disjoint union of a torus and a circle,
$(\mathbb{S}^1\times\mathbb{S}^1)\vee\mathbb{S}^1$. 
Whereas, at the local maximum $h=h_3$, 
${\cal S}_{h,j}=E\times\hat{\gamma}$, with $E$ an eight figure composed by $\{e_3\}$ and two
homoclinic connections, and ${\cal S}_{h,j}$  is homeomorphic to two tori glued by an equator.

For $j=0$,  \ie, $p_\varphi=0$, 
%Then, we discuss the structure of $H=h,p_\varphi=j$ when $j=0$. In this case 
the potential $V_0$ % $V(\theta)$ 
has no singularities and the trajectories can pass through $\theta=0$ 
and $\theta=\pi$, which are the singularities of the parametrization. 
\begin{figure*}[t]
  \centering
  \includegraphics[scale=0.9]{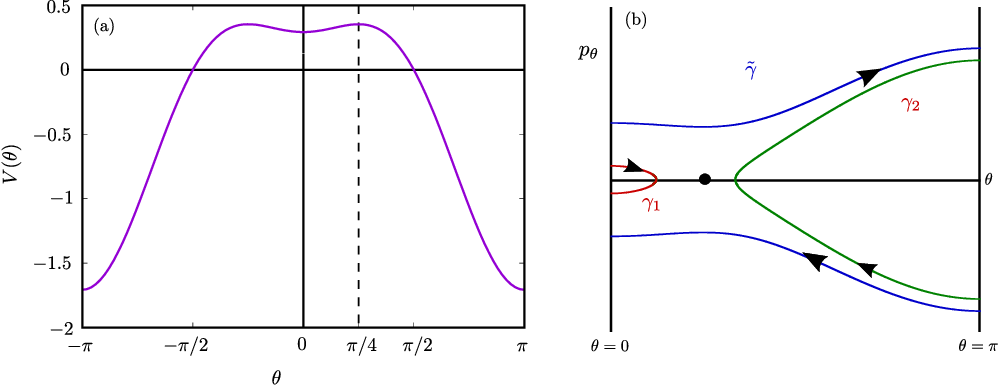}
\caption{(a) Potential $V(\theta)=\cos\theta-\frac{\sqrt{2}}{2}\cos^2\theta$ as a function of $\theta$, and (b) the orbits $\gamma_1$ and $\gamma_2$ with $h_1$ in the interval $V(0)<h_1<V(\pi/4)$, and $\tilde\gamma$ with $h_2$ satisfying $h_2>V(\pi/4)$  as a function of $\theta$. The orbit $\tilde\gamma$ 
corresponds to motions through a meridian of the sphere.}
\label{fig:fig3}
\end{figure*}
%% Primera figura de pagina 8bis
%\old{In conclusion, if $h$ is a regular value of $V_j$
%the set ${\cal S}_{h,j}$ %$\{H=h,p_\varphi=j\}$ 
%is composed by a finite family of invariant tori for both $j=0$ and $j\ne0$.
For this reason, it is convenient to consider the whole plane $(\theta,p_\theta)\in \mathbb{R}^2$ 
with the identifications $(\theta,p_\theta)\equiv(-\theta,-p_\theta)$, 
$(\theta+2\pi,p_\theta)\equiv(\theta,p_\theta)$. Since $\theta$ is the latitude on the sphere, 
this is consistent with the geometry of the problem. After reducing the phase space to the strip 
$[0,\pi]\times\mathbb{R}$, the orbits touching the vertical lines $\theta=0$ or $\theta=\pi$ will jump from $(\theta,p_\theta)$ to $(\theta,-p_\theta)$. For a given 
%\sout{a number $h$ which is a regular  value of $V$} 
regular value $h$ of $V$, there exists a finite number of closed orbits 
$\gamma_1,\ldots\gamma_r$ with energy $h$. As an example,
the closed curves  $\gamma_1$, $\gamma_2$ and $\tilde\gamma$    with $r=2$ and $r=1$, respectively, 
are presented in Fig.~\ref{fig:fig3}. The geometrical structures of 
the phase space $(\theta,p_\theta,\varphi,p_\varphi)$ are the invariant tori 
$\gamma_i\times\{(\varphi,0):\varphi\in[0,2\pi]\}$.

In conclusion, if $h$ is a regular value of $V_j$ the set $S_{h,j}$ is composed by 
a finite family of invariant tori for both $j=0$ and $j\ne 0$.

\subsection{The energy-momentum map}
\label{sec:energymomentum}
The energy-momentum map is defined as follows
\begin{equation*}
  \mathcal{EM}:M\rightarrow \mathbb{R}^2,(x,p)\rightarrow (H,p_\varphi),
\end{equation*}
 and associates an energy-momentum couple to each state. 
 The range of $\mathcal{EM}$ is
 \begin{equation*}
   \mathcal{EM}(M)=\left\{(h,j)\in\mathbb{R}^2:h\ge\min_{[0,\pi]} V_j\right\}.
 \end{equation*}
As expected, this map does not cover the whole $\mathbb{R}^2$ because  not all the 
 couples $(h,j)\in\mathbb{R}^2$ are admissible. 
A point $(h,j)\in \mathbb{R}^2$ is a regular value if the differential of $\mathcal{EM}$ is onto for
all the states $(x,p)\in M$ such that $\mathcal{EM}(x,p)=(h,j)$. Otherwise we say that $(h,j)$ is a
singular value. 
The validity of these definitions is demonstrated by the use of the coordinates
$(\theta,p_\theta,\varphi,p_\varphi)$, and the differential of $\mathcal{EM}$ given by
the matrix~\eqref{eq:sim}. This differential is onto wherever this matrix~\eqref{eq:sim} has rank two. 
As a consequence,  $(h,j)$, with  $j\ne0$, is a regular value of $\mathcal{EM}$ if and only if $h$ 
is a regular value of $V_j$, and the same conclusion holds for $j=0$.
The local surjectivity theorem~\cite{AMR} implies that every regular 
value must be in the interior of $\mathcal{EM}(M)$. 
Hence,  all points $(h,j)$ lying on the boundary of $\mathcal{EM}(M)$
are singular values. Additional singular values can appear in the interior of $\mathcal{EM}(M)$.

\subsection{Action-angle variables: local theory}
\label{sec:lth}
\begin{figure}[h]
  \centering
  \includegraphics[scale=0.9]{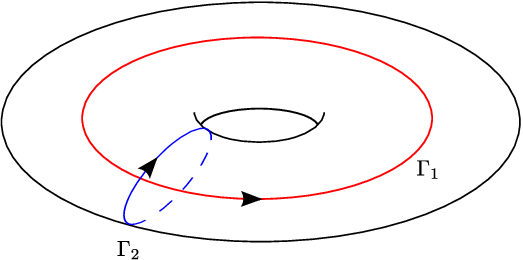}
  \caption{Oriented loops $\Gamma_1(h,j)$ and $\Gamma_2(h,j)$ in the invariant tori $\mathcal{T}=\mathcal{T}(h,j)$.} 
  \label{fig:fig4}
\end{figure}

Given  an invariant torus $\mathcal{T}_0$ such that the differential of $\mathcal{EM}$ has rank two over all
the points of $\mathcal{T}_0$, the theorem of Liouville-Arnold~\cite{Moser2006} 
allows for a change of variables on a neighborhood of $\mathcal{T}_0$
\begin{equation*}
  (x,p)\rightarrow(I_1,I_2,\varphi_1,\varphi_2)
\end{equation*}
with $\varphi_i=\varphi_i+2\pi$, and the form is given by 
\begin{equation*}
  \mathrm{d}\varphi\wedge\mathrm{d}p_\varphi+\mathrm{d}\theta\wedge\mathrm{d}p_\theta=\mathrm{d}\varphi_1\wedge\mathrm{d}I_1+\mathrm{d}\varphi_2\wedge\mathrm{d}I_2.
\end{equation*}
Now, the Hamiltonian only depends on the actions  $ H=H(I_1,I_2)$, and the
equations of motion become
\begin{equation}
  \label{eq:aa}
  \dot\varphi_i=\cfrac{\partial H}{\partial I_i}, \quad \dot I_i=0
\end{equation}
with $i=1,2$.
The phase space is foliated by invariant tori $I_1=c_1$, and  $I_2=c_2$, which 
% and we assume that  they 
can  be also labeled in terms of $(h,j)$, say $\mathcal{T}_0=\mathcal{T}_0(h,j)$. 
This is always the case in a neighborhood of $\mathcal{T}_0=\mathcal{T}(h_0,j_0)$.
The actions are constructed following  Arnol'd and  Novikov~\cite{Arnold1990a}. 
First, two oriented loops $\Gamma_1(h,j)$, and $\Gamma_2(h,j)$ are selected
 in $\mathcal{T}(h,j)$ to generate the first Homology group of the torus in~\autoref{fig:fig4}.
%% Primera figura de pagina 11
These loops must be chosen  so that they depend continuously on $(h,j)$. The actions
are computed via the integrals
\begin{equation}
  \label{eq:act}
  I_i=\cfrac{1}{2\pi}\int\limits_{\Gamma_i}\left(p_\theta\mathrm{d}\theta+p_\varphi\mathrm{d}\varphi\right).
\end{equation}

Let us consider a domain $G^+\subset\{j>0\}$, which is fibered by invariant tori and is small enough, so that the Liouville-Arnold theorem is applicable. 
 %where we can consider
Assume that $\gamma=\gamma(h,j)$ is  a closed orbit  of Eq.~\eqref{eq:re}, 
%\begin{equation*}
%\cfrac{1}{2}\dot \theta^2+V_j(\theta)=h,
%\end{equation*}
which depends analytically on $(h,j)$.
Assuming that $\gamma$ is the projection of $\mathcal{T}(h,j)$
on the plane $(\theta,p_\theta)$,
the loops are defined as 
\begin{equation*}
  \Gamma_1(h,j)=\gamma(h,j)\times\{(\pi,j)\},\quad \Gamma_2(h,j)=\{(\theta^*,p_\theta^*)\}\times\widehat{\gamma},
\end{equation*}
where $(\theta^*,p_\theta^*)$ is a point in $\gamma$ and
$\widehat{\gamma}=\widehat{\gamma}(j)=\{(\varphi,j):\varphi\in[0,2\pi]\}$.
Combining this definition with Eq.~\eqref{eq:act}, it yields 
\begin{equation}
  \label{eq:act2}
  \left\{
    \begin{array}{l}
      I_1=\cfrac{1}{2\pi}\int\limits_{\Gamma_1}p_\theta\mathrm{d}\theta=\cfrac{1}{\pi}\int_{\alpha_-}^{\alpha_+}\sqrt{2(h-V_j(\theta))}\mathrm{d}\theta,\\
I_2=\cfrac{1}{2\pi}\int\limits_{\Gamma_2}p_\varphi\mathrm{d}\varphi=j,
    \end{array}
  \right.
\end{equation}
where $\alpha_\pm=\alpha_\pm(h,j)$ are the points of $\gamma$ lying on the horizontal axis
$p_\theta=0$ satisfying $V_j(\alpha_\pm)=h$. The value of $\alpha_+$ will be $0$ (resp. $\pi$) when $\gamma$ does not intersect $p_\theta=0$ by the left (resp. by the right).
The action $I_1=I_1(h,j)$ is analytic on $G^+$. 
The actions can analogously be defined on the symmetric domain
$G^-\subset\{j<0\}$. 

It is interesting to discuss the behavior of the points % the numbers
$\alpha_\pm$ %=\alpha_\pm(h,j)$ 
as $j\rightarrow0$ and $h\rightarrow h_0$. 
 Assume that $\gamma(h,j)$ converges to a closed orbit $\gamma(h_0,0)$ where $h_0$ is a regular value of $V_0$ with $\min V_0< h_0< \max V_0$, then
 \begin{equation*}
  \lim_{(h,j)\rightarrow(h_0,0)}\alpha_\pm(h,j)=\alpha_\pm(h_0,0)
  \end{equation*}
with $0\le \alpha_-(h_0,0)<\alpha_+(h_0,0)\le \pi$.

For $h_0>\max V_0$, the orbit $\gamma(h,j)$ converges to the closed curve composed by the arcs $\tilde{\gamma}$ and two segments of the lines $\theta=0$ and $\theta=\pi$ (see Fig.~\ref{fig:fig3}). As a consequence
 \begin{equation*}
  \lim_{(h,j)\rightarrow(h_0,0)}\alpha_-(h,j)=0,\quad \lim_{(h,j)\rightarrow(h_0,0)}\alpha_+(h,j)=\pi.
\end{equation*}

\subsection{Action-angle variables: global aspects}
\label{sec:actionangle}
The action-angle variables are quite rigid, and this implies that their global definition requires a deeper analysis. 
Given a domain $D\subset M$ invariant under the  Hamiltonian flow, 
action-angle variables can be defined on $D$ if there is a symplectic diffeomorphism
\begin{equation*}
  \psi:D\subset M\rightarrow D_1\subset \mathbb{R}^2\times\mathbb{S}^1\times\mathbb{S}^1,\quad (x,p)\rightarrow (I_1,I_2,\varphi_1,\varphi_2)
\end{equation*}
with $H=H(I_1,I_2)$.
Given two systems of action-angle $\psi:D\rightarrow D_1$
and $\widetilde{\psi}:\widetilde{D}\rightarrow\widetilde{D}_1$, there exists a  $2\times 2$ matrix $A$,
with  constant integer entries and $\det A= 1$, and a constant vector $c\in \mathbb{R}^2$, 
such that the
actions in the two systems satisfy
\begin{equation}
  \label{eq:comp}
  \begin{pmatrix}
    I_1\\
    I_2
  \end{pmatrix}=A
  \begin{pmatrix}
    \widetilde{I_1}\\
    \widetilde{I_2}
  \end{pmatrix}+c
\end{equation}
on $D\cap \widetilde{D}$.
This is valid for any integrable Hamiltonian system such that the frequencies of the invariant tori,
%\begin{equation*}
  $\omega_i=\frac{\partial H}{\partial I_i}$ %\quad   \omega_2=\cfrac{\partial H}{\partial I_2}
%\end{equation*}
with $i=1,2$, are $\mathbb{Z}$-linearly independent and this holds for almost all tori~\cite{Moser2006}.
For the systems considered in this work, this is equivalent
to say that the periods of  $\theta(t)$ and  $\varphi(t)$, $T(h,j)$  in Eq.~\eqref{eq:f1}
and $\widehat{T}(h,j)$ in Eq.~\eqref{eq:f2},
 are not commensurable for most $(h,j)$. This is satisfied  because the ratio
$T(h,j)/\widehat{T}(h,j)$ is not constant. Indeed, if $(h,j)\rightarrow(h_0,0)$, with $h_0$ being a 
regular value of $V_0$ and $\min V_0< h_0<\max V_0$, then $T(h,j)$ tends to a positive 
number,  whereas $\widehat{T}(h,j)$ diverges to infinity. 
%\red{Here we are using the discussion at the end of  section \ref{sec:lth} and the formulas\eqref{eq:f1} and \eqref{eq:f2}.}

\section{Monodromy}
\label{sec:monodromy}
\begin{figure}[h]
  \centering
    \includegraphics[scale=0.7]{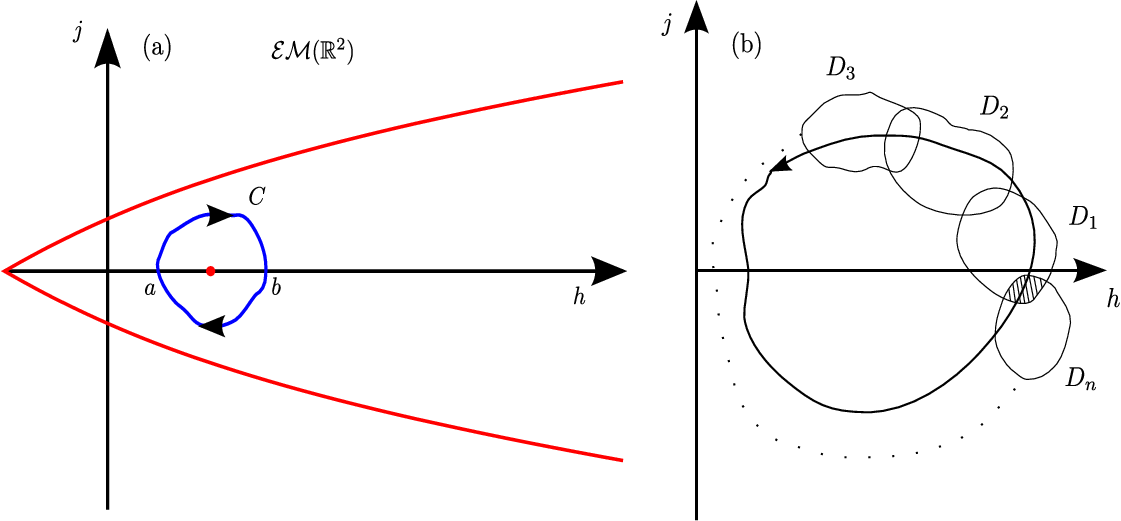}
\caption{(a) Circuit $C$ around a singular value in the line $j=0$ in the plane $(h,j)$ contained in the region of regular values of $\mathcal{EM}$ and (b) example of a family of domains $D_k$ covering the contour $C$.}
  \label{fig:fig5}
\end{figure}

In the plane $(h,j)$,  we consider a circuit $C$  contained in the region of regular values of
$\mathcal{EM}$, which crosses the line $j=0$ at two points $h=a$ and  $b$, and encircles a singular value lying on the line $j=0$, see~\autoref{fig:fig5}.
Associated to this circuit, there is a family  of invariant tori  $\mathcal{T}=\mathcal{T}(h,j)$, 
$(h,j)\in C$, depending continuously on $(h,j)$. This
circuit admits a system of action-angle variables defined in the domain $D$ if 
there is a sympletic diffeomorphism $\psi$ satisfying the conditions 
given in
section~\ref{sec:actionangle}, and the domain $D$ contains all invariant tori $\mathcal{T}(h,j)$. Note that it would be more precise to say that the family of invariant tori $\mathcal{T}$ admits these coordinates. The Hamiltonian system possesses monodromy if some circuit in $\mathcal{EM}$ does not admit action-angle coordinates~\cite{Duistermaat1980}.

%In the rest of the paper 
To derive a semi-theoretical method to detect  monodromy, it will be assumed that the family of invariant tori
$\mathcal{T}$ is of the first kind described in Section~\ref{sec:inv}. This means that for each $(h,j)\in C$ the invariant torus $\gamma(h,j)\times\hat\gamma$ is such that $h>\min V_j$ is a regular value of the modified potential and $h<\max V_0$ if $j=0$. In particular, it holds
\begin{equation}
    0<\alpha_-(h,j)<\alpha_+(h,j)<\pi,\qquad (h,j)\in C, \,\,j\ne 0.
\end{equation}
The remaining cases can be treated similarly. For each $(h,j)$ with $j\ne 0$, we define the function
\begin{equation}
    \label{eq:alpha}
    \chi(h,j)=-\cfrac{j}{\pi}\int_{\alpha_-}^{\alpha_+}\cfrac{1}{\sqrt{2(h-V_j(\theta))}}\cfrac{\mathrm{d}\theta}{\sin^2\theta}.
\end{equation}
The integrand has a singularity at $\theta=\alpha_\pm$, which is of the order of $|\theta-\alpha_\pm|^{-1/2}$, and, therefore, the integral is finite.

At the point $(a,0)$, which lies in the intersection of the circuit and the line $j=0$, we define
\begin{equation}
\label{eq:delta}
\Delta(a,0)=\displaystyle\lim_{\substack{
     (h,j)\to (a,0)\\
     j>0}
}2\chi(h,j).
\end{equation}
We will prove that this limit always exists and it takes integer values. 
 Analogously,  $\Delta(b,0)$ is defined at the point  $(b,0)$, which also takes
 integer values. 
 The criterion for the existence of monodromy along the circuit $C$ is 
\begin{equation}
    \label{eq:criterion_monodromy}
     \Delta(a,0)\ne\Delta(b,0).
\end{equation}

To justify these statements let us go back to 
the definition of action variables in~\autoref{eq:act2}. They are well defined in a small neighborhood of $C_+=C\cap\{j>0\}$ or $C_-=C\cap\{j<0\}$. The Jacobian matrix of $I=(I_1,I_2)$ can be computed in each of these two regions, 
\begin{equation*}
  DI(h,j)=  %D\vec{I}(h,j)=
  \begin{pmatrix}
    \beta & \chi\\
0 & 1
  \end{pmatrix}
\end{equation*}
with 
\begin{equation}
    \label{eq:beta}
    \beta=\beta(h,j)=\cfrac{1}{\pi}\int_{\alpha_-}^{\alpha_+}\cfrac{\mathrm{d}\theta}{\sqrt{2(h-V_j(\theta))}}.
\end{equation}
Note that there are not additional terms involving the derivatives of $\alpha_\pm(h,j)$ 
because $V(\alpha_\pm)=h$. Since $\beta>0$, $DI(h,j)$ has an inverse for $j\ne 0$, and it is straightforward to show that $\beta(h,-j)=\beta(h,j),\chi(h,-j)=-\chi(h,j)$ and 
\begin{eqnarray}
\label{eq:nilpo}
    DI(h,j)DI(h,-j)^{-1}&=&
  \begin{pmatrix}
    1 & 2\chi(h,j)\\
0 & 1
  \end{pmatrix}.
\end{eqnarray}

\begin{figure}[h]
    \centering
    \includegraphics[scale=0.8]{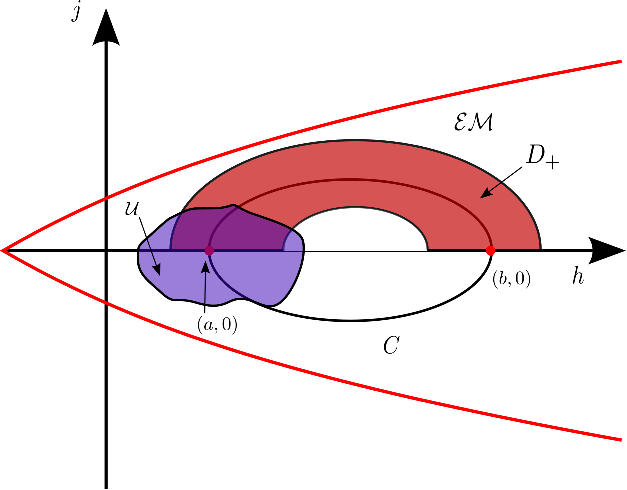}
    \caption{Domain $D_+$ in a neighbourhood of $C\cap\{j>0\}$, and 
    neighbourhood of $(a,0)$, $\mathcal{U}$, in which the actions  $\widetilde I_1$ and $\widetilde I_2$, are smooth functions of $(h,j)$.}
    %\jjalert{Needs an appropriate caption, that I feel unable to write}
    \label{fig:fig6}
\end{figure}

Let us now prove that the limit defining $\Delta(a,0)$ in~\autoref{eq:delta} really exists, and it is an integer number, which also holds for $\Delta(b,0)$.  
The Liouville-Arnold theorem can be applied at the invariant torus $\mathcal{T}(a,0)$. The local action-angle variables are defined on some open set $D_a\subset M$ with $\mathcal{T}(a,0)\subset D_a$. The continuity of the family of tori allows to find a neighborhood of $(a,0)$, $\mathcal{U}\subset\mathcal{EM}$, such that
\begin{equation}
    \mathcal{T}(h,j)\subset D_a\quad\text{if}\quad (h,j)\in\mathcal{U}.
\end{equation}
The corresponding actions, denoted by $\widetilde I_1$ and $\widetilde I_2$, are smooth functions of $(h,j)\in\mathcal{U}$. The actions $I_1$ and $I_2$, given by~\autoref{eq:act2}, are also smooth functions of $(h,j)$ but they are defined in a region of $\{j>0\}$. More precisely, in a neighborhood $D_+$ of $C\cap\{j>0\}$ 
as shown in~\autoref{fig:fig6}. 
According to~\autoref{eq:comp}, in the common region $D_+\cap \mathcal{U}$, it holds 
\begin{equation}
    DI(h,j)=A_+D\widetilde I(h,j),
\end{equation}
where $A_+$ is a matrix in the unimodular group, \ie, it has integer entries and $\det A_+=1$.
Since $\widetilde I(h,j)$ is
a smooth function in a neighborhood of $(a,0)$, the limit
\begin{equation}
    DI(a,0^+):=\lim\limits_{\substack{
    (h,j)\to (a,0)\\
         j>0 
}}DI(h,j)=A_+D\widetilde I(a,0)
\end{equation}
exists. Furthermore, since $\chi(h,j)$ is one of the entries of the matrix $DI(h,j)$, the limit $\Delta(a,0)$ in~\autoref{eq:delta} also exists.

Working on $\{j<0\}$ we obtain another limit $DI(a,0^-)$ with
\begin{equation}
    DI(a,0^-)=A_-D\widetilde I(a,0)
\end{equation}
being $A_-$ another unimodular matrix. In consequence,
\begin{equation}
\label{eq:daplusm}
    DI(a,0^+)DI(a,0^-)^{-1}=A_+A_-^{-1},
\end{equation}
and, from~\autoref{eq:nilpo},
\begin{equation}
\label{eq:daplusm_and2}
    DI(a,0^+)DI(a,0^-)^{-1}=
  \begin{pmatrix}
    1 & \Delta(a,0)\\
0 & 1
  \end{pmatrix}.
\end{equation}
Since the matrix $A_+A_-^{-1}$ is unimodular, the entry $\Delta(a,0)$ must be an integer.
The previous discussion is also applicable at the point $(b,0)$, obtaining that $\Delta(b,0)$ is also a  well defined integer.

\begin{figure}[h]
    \centering
    \includegraphics[scale=0.8]{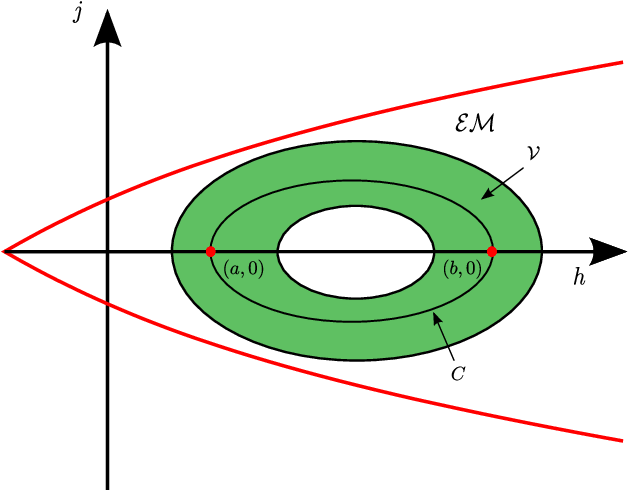}
    \caption{Neighborhood of $C$, $\mathcal{V}$,
    where the action variables $\mathcal{I}_i=\mathcal{I}_i(h,j)$, with
    $i=1,2$ are defined.}
    \label{fig:fig7}
\end{figure}

Let us now explain why $\Delta(a,0)\ne\Delta(b,0)$ implies monodromy. 
In the absence of it, there should exist a global system of action-angle coordinates. This means that there exists a diffeomorphism $\psi$, satisfying the conditions of Section~\ref{sec:lth}, defined on a domain $D$ containing the whole family of invariant tori, $\mathcal{T}(h,j)\subset D$ for each $(h,j)\in C$. The action variables associated to this system $\mathcal{I}_1$ and  $\mathcal{I}_2$ are smooth functions of $(h,j)$,  $\mathcal{I}_i=\mathcal{I}_i(h,j)$, and 
defined on some neighborhood of $C$, say $\mathcal{V}\subset \mathcal{EM}$, see~\autoref{fig:fig7}. 
By previous arguments, we know that there are unimodular matrices, again denoted by $A_\pm$, and it holds
\begin{eqnarray}
DI(h,j)&=&A_+D\mathcal{I}(h,j)\qquad\text{on}\quad \mathcal{V}\cap D_+,\\
DI(h,j)&=&A_-D\mathcal{I}(h,j)\qquad\text{on}\quad \mathcal{V}\cap D_-.
\end{eqnarray}
Then, it yields
\begin{equation}
DI(a,0^+)DI(a,0^-)^{-1}=\lim\limits_{\substack{
    (h,j)\to (a,0)\\
     j>0}
     }A_+D\mathcal{I}(h,j)
D\mathcal{I}(h,-j)^{-1}A_-^{-1}.=A_+A_-^{-1}.
\end{equation}
Since the same argument applies at the point $b$, obtaining
\begin{equation}
    DI(b,0^+)DI(b,0^-)^{-1}=A_+A_-^{-1},
\end{equation}
this implies that $\Delta(a,0)=\Delta(b,0)$. 
In conclusion, the existence of monodromy can be confirmed by 
numerically checking that
$\Delta(a,0)\ne\Delta(b,0)$, which means that  
a global system of action-angle coordinates cannot be defined~\cite{Duistermaat1980}.

%\section{Numerical study of the monodromy test}
\section{Numerical evaluation of monodromy}
\label{sec:numerical_tests}

This section is devoted to illustrate the validity of the monodromy test by applying it to several physical 
systems, which are known to present the phenomenon of monodromy. The generic potential for these systems  is given by
\begin{equation}
\label{eq:v_general}
    V(\theta)=-\omega\cos\theta-\eta\cos^2\theta-\lambda\cos^3\theta,
\end{equation}
with $\lambda,\eta,\omega\in \mathbb{R}$.

\begin{figure}[t]
  \centering
  \includegraphics[scale=0.8]{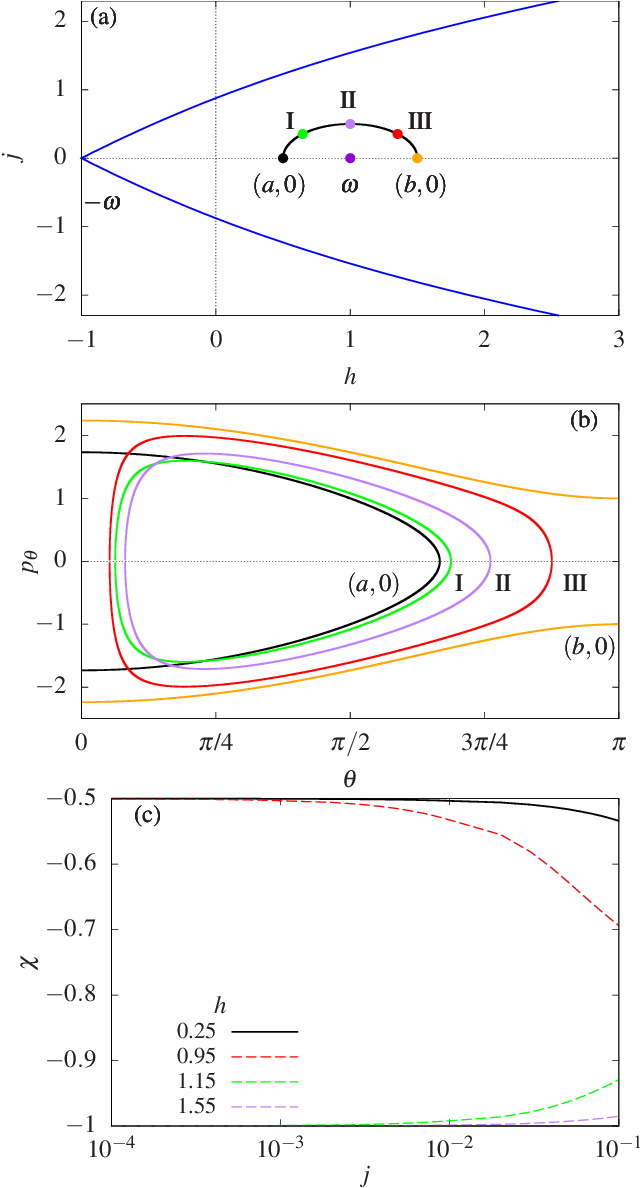}
  \caption{\label{fig:fig8} For the spherical pendulum with $\omega=1$, (a) the energy momentum plane $\mathcal{EM}$ and tori I=$(1-\frac{\sqrt{2}}{4},\frac{\sqrt{2}}{4} )$, II=$(1,\frac{1}{2} )$ and III=$(1+\frac{\sqrt{2}}{4},\frac{\sqrt{2}}{4} )$ (full points) along a path connecting $(a,0)=(\frac{1}{2},0)$ and $(b,0)=(\frac{3}{2},0)$ around the singular point $(0,\omega)$
  (b) projections of the tori on the plane $(\theta,p_\theta)$, and (c)
  $\chi(h,j)$ as a function of $j$ for several values of $h$.}
\end{figure}
As a first example, we consider the classical spherical pendulum,   $V(\theta)=-\omega\cos\theta$,
with $ \omega>0$ and $\lambda=\eta=0$, equivalent to $V(\theta)=\omega\cos\theta$.
This potential also describes the interaction of a polar rigid molecule with a static electric field parallel to 
the $Z$ axis of the laboratory fixed frame with $\omega=d \epsilon$, $d$ being the permanent electric dipole moment 
of the molecule and $\epsilon$ the  electric field strength~\cite{meyenn1970}.  
The equation~\eqref{eq:singularidad} for the singular points,  
with the change of coordinate  $x=\cos\theta$, reads
\begin{equation}
\label{eq:not_dif_static}
\omega(1-x^2)^2-j^2x=0.
\end{equation}
This equation has exactly one root lying in the interval $[-1,1]$ for $j\ne 0$, and 
the roots $\pm 1$ for $j=0$.
%\old{The  corresponding energy-momentum region $\mathcal{EM}$ is defined by the curve $\mu(j)$ and the isolated point$(0,\omega)$.}
The singularities on the energy-momentum region $\mathcal{EM}$ are the isolated point $(\omega,0)$ and the boundary, described by the curve of equation $h=\frac{j^2}{2(1-x(j)^2)}-\omega x(j)$ where $x(j)$ is the root of \autoref{eq:not_dif_static} lying in $[-1,1]$.
The $\mathcal{EM}$ map is presented in~\autoref{fig:fig8}~(a) where $\omega$ has been fixed to  $\omega=1$ without lost of generality.
Let us consider the family of invariant tori through a path in $\mathcal{EM}\cap \{j\ge 0\}$ connecting the points $(a,0)$ and $(b,0)$ with $-\omega<a<\omega<b$ indicated 
in~\autoref{fig:fig8}~(a). Specifically, we choose three tori along this 
path and present their projections on the plane $(\theta,p_\theta)$
in~\autoref{fig:fig8}~(b). As we observe, these trajectories smoothly connect the cases at each side of the singularity $(\omega,0)$, that is to say, open at $\theta=0$ and open at $\theta=0$ and $\pi$ for $(a,0)$ and $(b,0)$, respectively.

~\autoref{fig:fig8}~(c) shows $\chi(h,j)$ as a function of $j$  for several  values of $h$. 
In Appendix~\ref{sec:appen_numerics}, the numerical procedure to compute the integral $\chi(h,j)$ is explained. 
For $h>1$, it holds $\lim_{j\rightarrow0^+}\chi(h,j)=-1$, whereas for $h<1$, $\lim_{j\rightarrow0^+}\chi(h,j)=-1/2$. 
Thus, the inequality in~\autoref{eq:criterion_monodromy} is satisfied because
\begin{equation}
\label{eq:cond_monodromy_a_static}
  %\cfrac{1}{\pi}
\left[\Delta(b,0^+)-\Delta(a,0^+)\right] =-1,\quad 
\end{equation}
with $-1<a<1$ and $b>1$, which indicates the existence of monodromy.

%\begin{figure}[b]
%  \centering
%  \includegraphics[scale=0.8]{spherical_pendulum_static_w_1_em_aplus.eps}
%  \caption{For the spherical pendulum with $\omega=1$, (a) energy momentum plane, $\mathcal{EM}$, and (b) $\chi(h,j)$ for different values of $h$ as a function of $j$.}

%  \label{fig:em_static}
%\end{figure}

\begin{figure}[bt]
  \centering
  \includegraphics[scale=0.8]{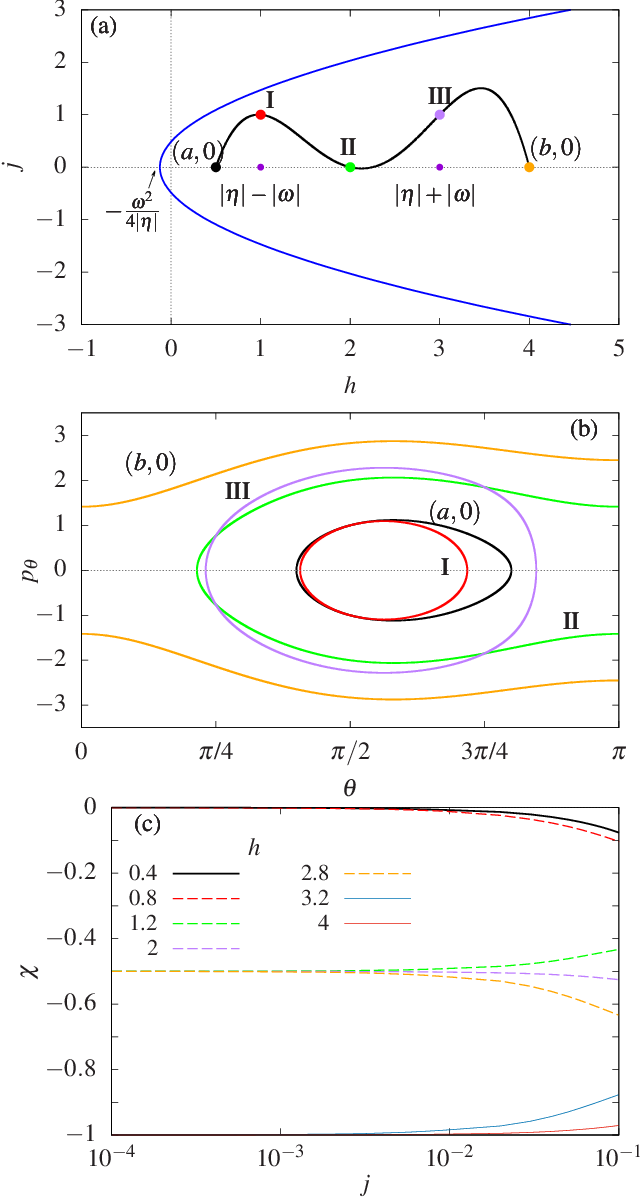}
  \caption{\label{fig:fig9}
  For the modified spherical pendulum with $\omega=-1$ and $\eta=-2$, (a) the energy momentum plane $\mathcal{EM}$ and 
  tori I=$(1,1)$, II=$(2,0)$ and III=$(3, 1)$ (full points) along a path connecting $(a,0)=(0.5,0)$ and $(b,0)=(4,0)$, 
  (b) projections of the tori on the plane $(\theta,p_\theta)$, and (c)
  $\chi(h,j)$ as a function of $j$ for several values of $h$.}
\end{figure}
The next system is a perturbed classical spherical pendulum
$V(\theta)=-\omega\cos\theta-\eta\cos^2\theta$, with $ \omega,\eta\lt 0$,  and $\omega\gt 2\eta$.
  Using again as coordinate $x=\cos\theta$ and  assuming $\theta\ne 0,\pi$,
 the singular-point  equation~\eqref{eq:singularidad}  reads
  \begin{equation}
  \label{eq:cond_monodromy_static_laser}
(\omega+2\eta x)(1-x^2)^2-j^2x=0.
\end{equation}
The computations show that this equation possesses only one solution for $j\ne 0$ lying in [-1,1], whereas  for $j=0$,  there are
three roots
lying in [-1,1] with values $\pm 1,-\frac{\omega}{2\eta}$. Thus, the singularities on $\mathcal{EM}$ for this potential are two isolated points and the boundary, with equation $h=\frac{j^2}{2(1-x(j)^2)}-\omega x(j)-\eta x(j)^2$, where $x(j)\in [-1,1]$ is the root of~\autoref{eq:cond_monodromy_static_laser}. Specifically, the singular points for $j=0$ are $-\omega-\eta=|\eta|+|\omega|$, $\omega-\eta=|\eta|-|\omega|$, and $\frac{\omega^2}{4\eta}=-\frac{\omega^2}{4|\eta|}$, which is included in the boundary. The $\mathcal{EM}$ map is  presented in Fig.~\ref{fig:fig9}~(a).
The points $|\eta|\pm|\omega|$ are the candidates to be the monodromy points, which are
 $1$ and $3$ for $\omega=-1$ and $\eta=-2$. 
Monodromy is illustrated along a path connecting three tori lying on the $j=0$ axis,
 see \autoref{fig:fig9}~(a). The projections on
$(p_\theta,\theta)$ plane of  the three tori I, II and III are closed, 
open on one side and open on both sides, respectively, as shown in~\autoref{fig:fig9}~(b). 
For several values of the total energy $h$, Fig.~\ref{fig:fig9}~(c) illustrates the evolution of $\chi(h,j)$
as $j$ decreases.  
It holds $\lim_{j\rightarrow 0^+}\chi(h,j)=0$ for $0<h<1$, 
 $\lim_{j\rightarrow 0^+}\chi(h,j)=-0.5$ for $1<h<3$, and  $\lim_{j\rightarrow 0^+}\chi(h,j)=-1$ for
$h>3 $. As  a consequence, for a closed trajectory containing  one or
two monodromy points, the inequality~\eqref{eq:criterion_monodromy} is satisfied, and, therefore, monodromy exists.

\begin{figure}[t]
  \centering
   \includegraphics[scale=0.8]{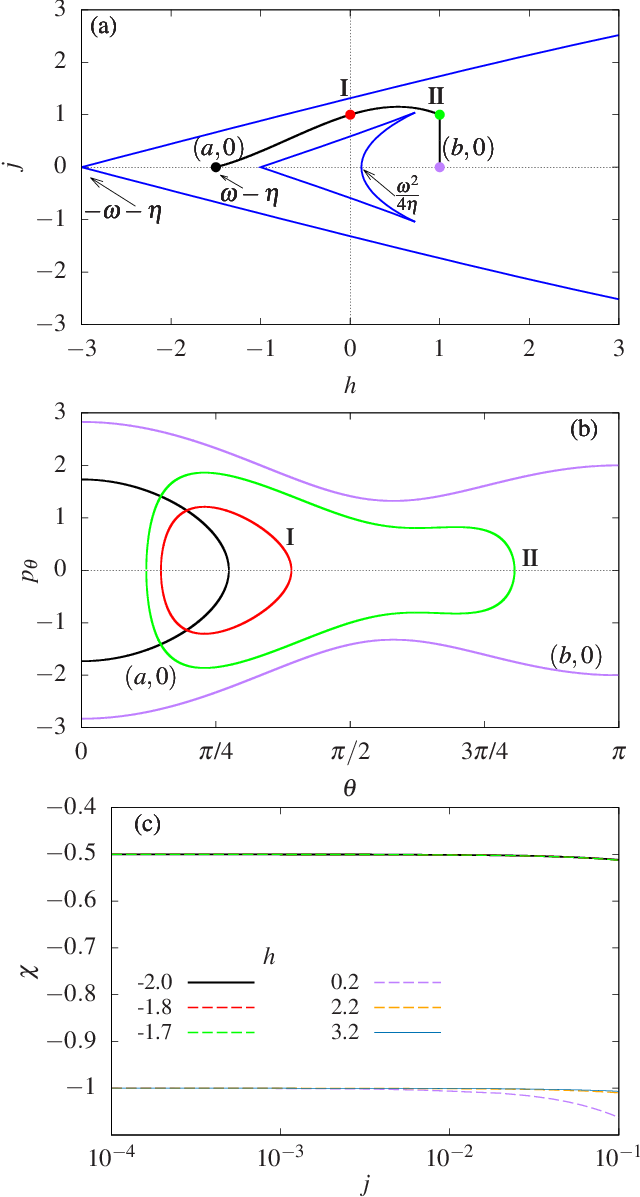}
\caption{\label{fig:fig10}
For the modified spherical pendulum with  $\omega=1$ and $\eta=2$, (a) the energy momentum plane $\mathcal{EM}$ and tori I=$(0,1)$ and II=$(1,1)$ (full points) along a path connecting $(a,0)=(-1.5,0)$ and $(b,0)=(1,0)$, 
  (b) projections of the tori on the plane $(\theta,p_\theta)$, and (c)
  $\chi(h,j)$ as a function of $j$ for several values of $h$.}
\end{figure}

By considering the parameters $\omega,\eta>0$,  
this potential  $V(\theta)=-\omega\cos\theta-\eta\cos^2\theta$ describes the interaction of a
polar rigid molecule with a static electric field and a non-resonant laser field~\cite{friedrich:jcp111,nielsen:prl2012},
both parallel to the $Z$ axis, with $\omega=d \epsilon$ and $\eta=I \Delta \alpha/2c\epsilon_0$, where
$\Delta\alpha$ is the polarizability anisotropy, $I$ the laser intensity, $c$ the speed of light and $\epsilon_0$ 
the dielectric constant. 
For $\omega<2\eta$, 
the singular-point equation~\eqref{eq:singularidad}, with   $x=\cos\theta$, is given by
\begin{equation}
  \label{eq:cond_monodromy_static_laser_neg}
(\omega+2\eta x)(1-x^2)^2-j^2x=0.
\end{equation}
For this potential, the singularities are $-\eta-\omega$, $-\eta+\omega$ and $\omega^2/4\eta$; whereas \autoref{eq:cond_monodromy_static_laser_neg} has one, two or three roots for $j\ne 0$. Thus, the set of regular values of~~$\mathcal{EM}$ has two connected components, as illustrated in Fig.~\ref{fig:fig10}~(a). 
For the invariant tori in Fig.~\ref{fig:fig10}~(a),
we present in Fig.~\ref{fig:fig10}~(b) 
their projections on the $(\theta,p_\theta)$ plane.
For $\omega=1$, $\eta=2$, and several values of $h$, the numerical results for $\chi(h,j)$
are presented as a function of $j$ in Fig.~\ref{fig:fig10}~(c). 
The limits $\lim_{j\rightarrow 0^+}\chi(h>\frac{\omega^2}{4\eta},j)$ and
$\lim_{j\rightarrow 0^+}\chi(h<\omega-\eta,j)$,  taken from the right or left side 
of the inner region, respectively, are different confirming the existence of monodromy in this system.

As a last example, we consider the general potential~\eqref{eq:v_general}, which represents the interaction of
a diatomic polar molecule in a two-color non-resonant laser field~\cite{oda2010}. The parameters
are  $\eta=\frac{1}{4}\Delta\alpha(\epsilon_1^2+\epsilon_2^2)$, $\lambda=\frac{1}{8}\Delta\beta\epsilon_1^2\epsilon_2\cos(2\delta_2-\delta_1)$, and  $\omega=\frac{3}{8}\beta_\perp\epsilon_1^2\epsilon_2\cos(2\delta_2-\delta_1)$, 
with 
$\beta_\bot$ and  $\Delta\beta$ being the hyperpolarizability perpendicular component and anisotropy, respectively,
$\epsilon_{1,2}$ and $\delta_{1,2}$ are the field strengths and phases of the two  components of
the laser field, respectively.
 The condition Eq.~\eqref{eq:singularidad} reads
\begin{equation}
    \label{eq:singularidad_twocolor}
    \left[2\eta x +3\lambda x^2+\omega\right](1-x^2)^2-j^2 x=0,
\end{equation}
being $x=\cos\theta$. 
%We now discuss several cases.
%\subsubsection{$\eta,\lambda,\,\omega>0$}
For $j=0$, the regularization condition~\eqref{eq:singularidad_twocolor}  
is fulfilled for $x=\pm 1$ and $x=\frac{1}{3\lambda}\left(-\eta\pm\sqrt{\eta^2-3\lambda\omega}\right)$. 
Thus, for $3\lambda\omega\gt \eta^2$, 
we encounter the roots $x=\pm 1$,
with energy values $h=\mp\lambda-\eta\mp\omega$, respectively. 
In addition to these two roots, a third one, $x=-\frac{\eta}{3\lambda}$, 
appears for 
$3\lambda\omega= \eta^2$,
and two more, $x=\frac{1}{3\lambda}\left(-\eta\pm\sqrt{\eta^2-3\lambda\omega}\right)$,  
for $3\lambda\omega\lt \eta^2$.
Here, we are assuming that the parameters 
$\omega$, $\eta$ and $\lambda$ satisfy the appropriate conditions so that 
the roots are in the interval $x\in [-1,1]$.

Due to the complexity of the roots, we only present the results for the 
parameters $\lambda=-1.1$, $\eta=-0.2$ and $\omega=1.2$.
The roots are $x=\pm 1$, $x\approx 0.5455$ and $x\approx -0.6667$, having 
the Hamiltonian the values $h=0.1$, $0.3$, $-0.4165$ and $0.5630$, respectively. 
 \begin{figure}[h]
    \centering
    \includegraphics[scale=0.8]{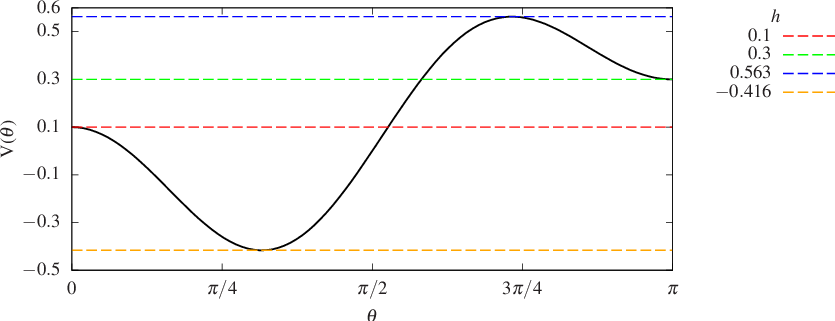}
    \caption{\label{fig:fig11}
   Interaction potential $V(\theta)=1.1\cos^3\theta+0.2\cos^2\theta-1.2\cos\theta$, the horizontal lines represent the energies at the singular points at $j=0$.}
\end{figure}
To illustrate this system, Fig.~\ref{fig:fig11} shows the interaction potential for $j=0$.
For $h\le -0.4165$, the system does not present orbits, since the kinetic energy cannot be larger than $0$. 
A unique type of orbit is found for  $-04165\lt h\lt 0.1$, in this case the values of  $\theta$ are restricted to the left well. 
By increasing the energy to $0.1\lt h\lt 0.3$, there is still one type of orbit, however, the system is able to access $\theta=0$.
Due to the potential barrier for $0.3\lt h \lt 0.563$, there are two regions in the phase space containing 
orbits.  If the energy of the system is larger than the potential maximum located 
around $3\pi/4$, \ie,  $h\gt 0.563$, the orbits cover the whole space.
 
\begin{figure}
    \centering
    \includegraphics[scale=0.8]{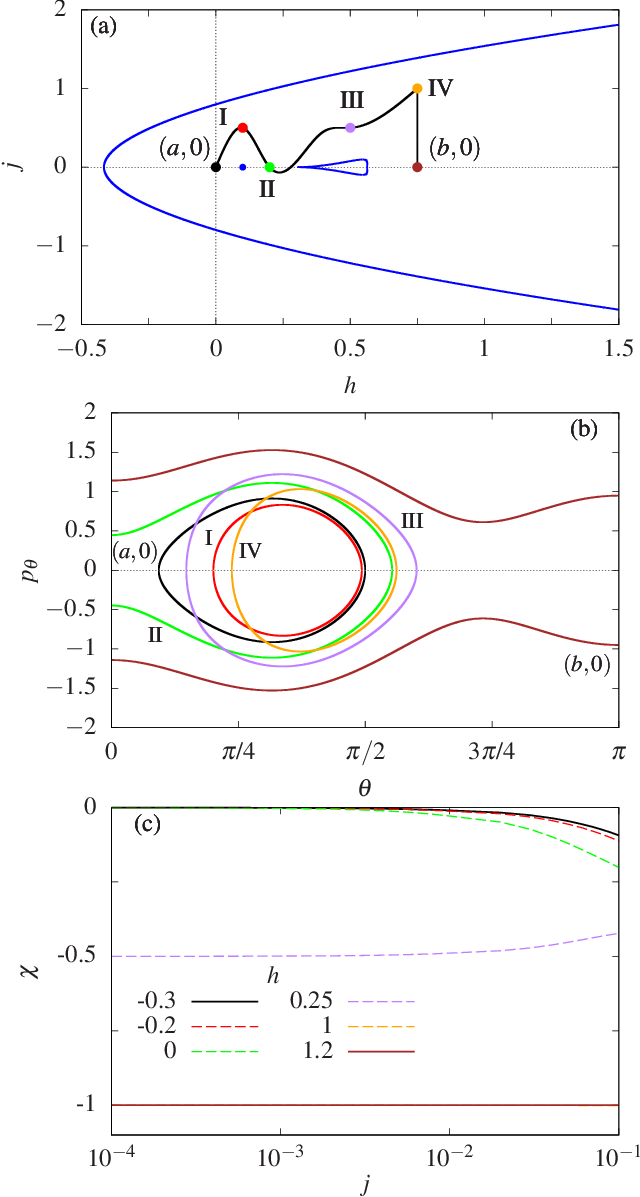}
    \caption{\label{fig:fig12} 
For the modified spherical pendulum with  $\omega=1.2$, $\eta=-0.2$ and $\lambda=-1.1$, (a) the energy momentum plane $\mathcal{EM}$ and 
tori I=$(0.1,5)$, II=$(0.2,0)$, III=$(0.5,0.5)$ and IV=$(0.75,1)$ (full points) along a path connecting $(a,0)=(0,0)$ and $(b,0)=(0.75,0)$, 
  (b) projections of these tori on the plane $(\theta,p_\theta)$, and (c)
  $\chi(h,j)$ as a function of $j$ for several values of $h$.}
\end{figure}

The corresponding $\mathcal{EM}$-map is presented 
in Fig.~\ref{fig:fig12}~(a), where
an irregular point and an irregular curve are observed. 
An example path around these irregular point and curve is also shown in this figure. 
The  projections on the $(\theta,p_\theta)$ plane of these invariant tori are 
plotted in Fig.~\ref{fig:fig12}~(b), where we observe all kind of trajectories. 
As in the previous cases, the limit $\lim_{j\rightarrow0^+}\chi(h,j)$,
presented in Fig.~\ref{fig:fig12}~(c),
depends on the value of $h$. We obtain
$\lim_{j\rightarrow0^+ }\chi(h,j)=-1,-0.5$ and $0$, for
$-0.4165\lt h \lt 0.1$, $0.1 \lt h \lt 0.3$ and $h\gt 0.563$, respectively, which
proves the existence of monodromy.

\section{Conclusions}
\label{sec:conclusions}
We have considered a particle whose motion is constrained to the unit sphere
and governed by  an external potential having azimuthal symmetry. 
For this system, the topology
and the local and global angle-action  variables are explored in detail. 
The monodromy  in a  system appears if the angle-action 
variables are impossible to define globally. In this work, we have presented a monodromy test to identify and 
characterize numerically the monodromy, based on the behavior of the closed orbits in the neighborhood of singular points or regions of the $\mathcal{EM}$ plane. 
This monodromy test can be performed systematically and does not require a deep mathematical knowledge of the equations of motion of the system. Furthermore, it can be easily implemented for different potentials to efficiently characterize the monodromy of a wide variety of systems. 
 The validity of this monodromy test has been illustrated in the vecinity of the singular points in the $(h,j)$ plane for several systems with azymuthal symmetry.

An interesting extension of this study covers the case of integrable systems with more degrees of freedom. 
It could also be of interest to extend these ideas to non-fully integrable systems. Of particular interests is the case of an rigid rotor interacting with an external field, such as the gravitational field. This kind of system is also interesting in molecular physics to model, for example, polyatomic molecules in combined parallel electric and nonresonant laser 
fields coupled to the polarizability~\cite{nielsen:prl2012,omiste:pra2012,Omiste2013} or THz laser fields coupled to the molecular hyperpolarizability~\cite{Coudert2017}. 

%For these potentials, the present methodology should be extended to account for more than one degree of freedom, and  the $\mathcal{EM}$ space has to be expanded beyond the plane.

\appendix
\section{The symplectic structure}
\label{sec:symplectic}
The space $\mathbb{R}^3\times \mathbb{R}^3$ is endowed with the 2-form
\begin{equation*}
  \omega=\sum_{i=1}^3\mathrm{d}x_i\wedge \mathrm{d}y_i.
\end{equation*}
The tangent bundle of the unit sphere
\begin{equation*}
  M=T\mathbb{S}^2=\{(x,p)\in\mathbb{R}^3\times\mathbb{R}^3:\parallel x\parallel=1,\langle x,p\rangle =0\}
\end{equation*}
is a symplectic submanifold. This is automatic since the Poisson bracket of the constraints never
vanishes~\cite{Moser2006}. Notice that if $F_1(x,y)=\parallel x\parallel^2-1$, $F_2(x,y)=\langle x,y\rangle$, then
\begin{equation*}
  \left\{F_1,F_2\right\}=\sum_{i=1}^3\left\{\cfrac{\partial F_1}{\partial x_i}\cfrac{\partial F_2}{\partial y_i}-\cfrac{\partial F_1}{\partial y_i}\cfrac{\partial F_2}{\partial x_i}\right\}=2.
\end{equation*}
In $M$ we introduce the Lagrangian coordinates
\begin{equation*}
  \left\{
    \begin{array}{l}
      x_1=\sin\theta\cos\varphi,\quad x_2=\sin\theta\sin\varphi,\quad x_3=\cos\theta\\
      y_1=-\dot\varphi\sin\theta\sin\varphi+\dot\theta\cos\theta\cos\varphi\\
      y_2=\dot\varphi\sin\theta\cos\varphi+\dot\theta\cos\theta\sin\varphi\\
      y_3=-\dot\theta\sin\theta
    \end{array}
    \right.
\end{equation*}
with $\theta\in ]0,\pi[$ and $\varphi\equiv \varphi+2\pi$.

The tangent planes at the north and south poles ($\theta=0$ and $\theta=\pi$) are not covered by
this chart. This will make some of our computations incomplete. The reader can easily fulfill the
remaining details by using alternatives charts including these planes.
After some computations it can be seen that the form induced by $\omega$ on $M$ is expressed as
\begin{equation*}
  \omega_M=\mathrm{d}\varphi\wedge\mathrm{d}p_\varphi+\mathrm{d}\theta\wedge \mathrm{d} p_\theta
\end{equation*}
with $p_\theta=\dot\theta$ and $p_\varphi=\dot\varphi\sin^2\theta$.
We will employ the symplectic coordinates
\begin{equation*}
  q=
  \begin{pmatrix}
    \theta\\
    \varphi
  \end{pmatrix},\qquad
p=
  \begin{pmatrix}
    p_\theta\\
    p_\varphi
  \end{pmatrix}.
\end{equation*}

\section{The numerical procedure}
\label{sec:appen_numerics}
The monodromy test requires the computation of the limits  $\lim_{j\rightarrow 0^\pm}\chi(h,j)$ in~\autoref{eq:alpha}
involving  the integral in Eq.~\eqref{eq:delta}, which must be done numerically since, in general, it 
is not solvable analytically. This integral~\eqref{eq:alpha} is improper because it is finite although the integrand  
diverges in the upper and lower limits, \ie, $\sqrt{h-V_j(\alpha_{\pm})}=0$. 
This fact constitutes an obstacle for the numerical integration, which is bypassed following
the procedure described in Ref.~\onlinecite{Press1992}. 
First,  the upper and lower limits, $\alpha_\pm$, 
%which are the turning points if $\cos\alpha_+\le 1$ and $\cos\alpha_-\ge -1$, respectively,
are obtained  by solving the equation $\sqrt{h-V_j(\alpha_{\pm})}=0$. 
The integral~\eqref{eq:alpha} is solved using the change of variables $x=\cos\theta$.
The  singularities in the upper and lower limits are treated as follows:
i) if the integrand $f(x)$ diverges as $(x-b)^{-\gamma}$, with $0\le\gamma<1$ and $b$ being the upper limit, 
 the singularities are removed by changing the  variable $x=b-t^{\frac{1}{1-\gamma}}$ leading  to
    \begin{equation}
        \label{eq:lim_b}
        \int_a^bf(x)\mathrm{d}x=\cfrac{1}{1-\gamma}\int_0^{{b-a}^{1-\gamma}}t^{\frac{\gamma}{1-\gamma}}f\left(b-t^{\frac{1}{1-\gamma}}\right); %\qquad (b>a),
    \end{equation}
ii) if $f(x)$ diverges as $(x-a)^{-\gamma}$, with $0\le\gamma<1$ and $a$ being the lower limit, the change of variable $x=a+t^{\frac{1}{1-\gamma}}$ results on
    \begin{equation}
    \label{eq:lim_a}
        \int_a^bf(x)\mathrm{d}x=\cfrac{1}{1-\gamma}\int_0^{{b-a}^{1-\gamma}}t^{\frac{\gamma}{1-\gamma}}f\left(a+t^{\frac{1}{1-\gamma}}\right); %\qquad (b>a).
    \end{equation}    
and iii) if $f(x)$ possesses singularities in both limits, as it occurs in the examples analyzed here,
the integral is split and computed in the intervals $a\le x\le c$ and $c\le x\le b$,  where $c\in(a,b)$ and $f(c)$ is 
well defined. Finally, the integrals~\eqref{eq:lim_b}, and~\eqref{eq:lim_a} are evaluated~\cite{Press1992}. 

\begin{acknowledgments}
J.J.O. acknowledges the funding from Juan de la Cierva - Incorporaci\'on program granted by Ministerio de Ciencia e Innovaci\'on (Spain) and Project PID2019-106732GB-I00 (MINECO).
R.G.F. acknowledges  financial support by the Spanish Project No. FIS2017-89349-P  (MINECO), 
and by the Andalusian research  group FQM-207. 
This study has been partially  financed by the Consejer\'{\i}a de 
Conocimiento, Investigaci\'on y Universidad, Junta de  Andaluc\'{\i}a and European Regional Development 
Fund (ERDF), Ref. SOMM17/6105/UGR.  R.G.F. completed some of this work as a Fulbright fellow at ITAMP at Harvard 
University.

\end{acknowledgments}

\bibliographystyle{apsrev4-1}
%\bibliography{Monodromy}
 
%

\end{document}